\documentclass[prd,aps,twocolumn]{revtex4}
\usepackage{epsfig}

\newcommand{\dd}{{\rm d}}

\def\mR{{\cal R}}

\def\dsl{\chi}
\def\sigdsl{\sigma_{\chi}}

\begin{document}
\title{Inflationary models inducing non-Gaussian metric fluctuations}

\author{Francis Bernardeau \vskip 0.1cm}
\affiliation{Service de Physique Th{\'e}orique,
         CEA/DSM/SPhT, Unit{\'e} de recherche associ{\'e}e au CNRS, CEA/Saclay
         91191 Gif-sur-Yvette c{\'e}dex}
\author{Jean--Philippe Uzan \vskip 0.1cm}
\affiliation{Laboratoire de Physique Th{\'e}orique, CNRS--UMR 8627,
         B{\^a}t. 210, Universit{\'e} Paris XI, F--91405 Orsay Cedex,
         France,\\
         Institut d'Astrophysique de Paris, GReCO,
        CNRS-FRE 2435, 98 bis, Bd Arago, 75014 Paris, France.
\vskip 0.15cm}
\date{\today}

\begin{abstract}
We construct explicit models of multi-field inflation in which the
primordial metric fluctuations do not necessarily obey Gaussian
statistics. These models are realizations of mechanisms in which
non-Gaussianity is first generated by a light scalar field and
then transferred into curvature fluctuations. The probability
distribution functions of the metric perturbation at the end of
inflation are computed. This provides a guideline for designing
strategies to search for non-Gaussian signals in future CMB and
large scale structure surveys.
\end{abstract}
\pacs{{ \bf PACS numbers:} } \vskip2pc

\maketitle

Inflation generically predicts Gaussian initial metric
fluctuations with an almost scale invariant power
spectrum~\cite{inflation}. It is interesting to determine how
general this property is.  With the advent of large scale
structure and CMB surveys it will be possible to test in details
whether the initial conditions were Gaussian or not and in the
latter case how they can be characterized.  At the moment, no
non-Gaussian signal has been detected in CMB data~\cite{cmb} but
the number of modes that can be probed is still small. In
large-scale structure surveys the number of independent modes that
are observed is large but the difficulty arises from the
non-linear gravitational dynamics~\cite{revue} which can shadow
the primordial non-Gaussianity~\cite{nous}. Having at our disposal
models of inflation in which non-Gaussian adiabatic metric
fluctuations are generated can then serve as a guideline for
designing strategies for detecting primordial non-Gaussianities.

In single field inflation, the slow-roll conditions, if valid
throughout the period during which the seeds of the large scale
structures are generated, prevent the generation of observable
primordial non-Gaussianities. The reason is that the potential
needs to be both flat enough for the fluctuations to develop and
steep enough for the non-linearity to be significant. Note however
that starting from a non-vacuum initial state~\cite{nonvide} or
allowing for sharp features in the potential~\cite{sharpfeature}
could lead to primordial non-Gaussianities. In those cases these
features will be localized in a narrow band of wavelengths and
will affect the shape of the density fluctuation power spectrum.

The situation is richer when more than one light scalar field are
present during inflation. In multi-field inflation models, a mixture
of adiabatic and isocurvature fluctuations~\cite{adia_iso}, that can
be correlated or not~\cite{correle}, are generically produced. It
opens the door to the generation of primordial
non-Gaussianity~\cite{multi_ng} because the non-linear couplings can
be much stronger in the isocurvature direction than in the (inflaton)
adiabatic direction~\cite{multi_ng,multi_ng2}.  When the fields are
coupled, exchange between isocurvature and adiabatic modes can be
important~\cite{bartolo} which makes it possible to generate
non-Gaussianity on a large band of wavelengths while preserving an
adiabatic slow-roll type primordial power spectrum~\cite{ub02}.

In a previous study~\cite{ub02}, we have presented the building
blocks for such models. It is based on the generation of
non-Gaussian isocurvature fluctuations which are subsequently
transferred to the adiabatic mode due to a bend in the classical
inflaton trajectory. The necessary ingredients for such models are
(i) the existence of a light scalar field that has non-linear
coupling, typically quartic, and (ii) a coupling term in the
potential leading to isocurvature-adiabatic mode mixing. This
latter mechanism can be given a simple geometrical interpretation.
On super Hubble scales, curvature fluctuations are induced by
fluctuations of the total expansion in different parts of the
universe. The relative duration of their inflationary phase can be
affected if inflaton trajectory is bent. Fluctuations in the
isocurvature modes generate a bundle of parallel trajectories
along which inflation lasts longer or shorter depending on whether
they lie in the outer or inner part of the bent (see Fig.~1 of
Ref.~\cite{ub02}).

In Ref.~\cite{ub02}, we characterized the expected statistical
properties of the metric fluctuations that were shown to be the
superposition of a Gaussian and a non-Gaussian contributions of
the same variance. The relative weight of the two contributions is
related to the bending angle in field space. We explicitly
computed the probability distribution function (PDF) of the
non-Gaussian contribution which appears to be described by a
single new parameter, $\nu_3$ [see Eq.~(\ref{pdfapprox}) below].
These results were obtained without the use of any explicit model.
This {\it letter} is dedicated to the building of this kind of
inflationary models involving either two or three fields. The
shape of their PDF will be computed and compared to our previous
analytical expression~\cite{ub02}.

In general the curvature fluctuations, $\mR$, can be computed on
superhorizon scales as~\cite{LiddleLythPR},
\begin{equation}\label{curv}
  \mR= H\delta t.
\end{equation}
Isocurvature fluctuations can then induce curvature fluctuations
only if in some way or another they can change the relative total
number of $e$-fold.


This can actually be obtained in a very simple way in hybrid type
model of inflation with three fields. In that case, one field,
$\phi$, is the inflaton; the second field is a light scalar,
$\chi$, with quartic coupling $\lambda\le 1$ and the third field,
$\sigma$, is coupled to the two others so that the end of
inflation is triggered when $\sigma$ undergoes a phase transition.
To be more explicit, let us examine the following model
\begin{eqnarray}\label{threefieldmodel}
 V(\phi,\chi,\sigma)&=&\frac{1}{2}m^2\,\phi^2
        +\frac{\lambda}{4!}\chi^4+
        \frac{\mu}{2}\left(\sigma^2-\sigma_0^2\right)^2
        \nonumber\\
  &&+\frac{g}{2}\sigma^2\left(\phi\cos\alpha+\chi\sin\alpha\right)^2
\end{eqnarray}
where $\sigma_0$ is the final vev of $\sigma$ and $\alpha$ is the
mixing angle between $\phi$ and $\chi$ in their coupling to
$\sigma$.

Inflation ends when the effective mass of $\sigma$ vanishes, that
is when
\begin{equation}\label{effectivemass}
  g\left(\phi\cos\alpha+\chi\sin\alpha\right)^2-2\mu\sigma_0^2=0.
\end{equation}
The value of $\phi$ at the end of inflation is
\begin{equation}\label{phiend}
  \phi_{\rm
  end}\equiv\frac{\pm\sqrt{2\mu/g}\,\sigma_0-\chi\sin\alpha}{\cos\alpha}.
\end{equation}
For $\phi>\phi_{\rm end}$, $\sigma=0$ and the two fields evolve
independently: $\phi$ drives the inflation while $\chi$ develops
non-Gaussianity. The amount of non-Gaussianity of $\chi$ then
depends only on $\lambda$ and on the total number of $e$-fold
between horizon crossing and the end of inflation.

When $\alpha$ is non-zero, fluctuations of $\chi$ induce curvature
fluctuations because they change the time at which the phase
transition occurs. Thus the $\chi$-induced curvature fluctuations
read (assuming $H$ is basically constant during the inflationary
period),
\begin{equation}\label{chiinducedR}
  \mR\simeq H \delta t_{\rm
  end}\simeq-\frac{3H^2}{V_{,\phi}}\,\frac{\sin \alpha}{\cos \alpha}\,
  \chi.
\end{equation}

\begin{figure}
  \centerline{ \psfig{figure=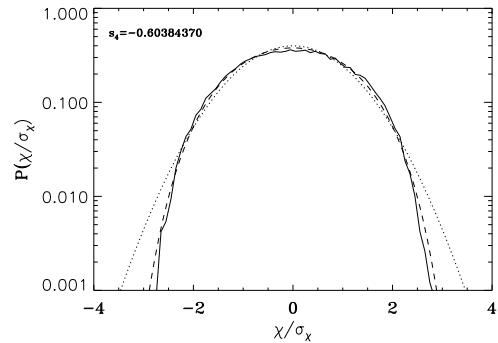,width=7cm}}
\caption{The shape of the probability distribution function of the
isocurvature modes in case of a quartic coupling,
\ref{threefieldmodel}, (solid line) compared to the analytic PDF of
Eq.~(\ref{pdfapprox}) (dashed line) and to Gaussian distribution (dotted
line). In this example, $\lambda=10^{-2}$ and $N_e=50$.} \label{chipdf}
\end{figure}

We present on Fig.~\ref{chipdf} the result of a numerical
investigation in which the super-Hubble evolution of the $\chi$
modes are explicitly computed. The Klein-Gordon equations for the
two fields and the Friedmann equations are integrated for a set of
initial conditions leading to a bundle of classical trajectories.
The parameters of this example were taken to be $\lambda=10^{-2}$
and an evolution for a number of $e$-folds equal to $N_e=50$. The
shape of the PDF is compared to our analytical
expression~\cite{ub02},
\begin{eqnarray}\label{pdfapprox}
  P(\dsl)\dd\dsl&=&
  \sqrt{\frac{1}{2\,\pi}\left|\frac{1-\dsl^2\nu_3}{(1+\dsl^2\,\nu_3/3)^3}\right|}
  \times\nonumber\\
  &&\exp\left[-\frac{3\,\dsl^2}{(6+2\,\dsl^2\,\nu_3)\sigdsl^2}\right]
  \frac{\dd\dsl}{\sigdsl}.
\end{eqnarray}
which depends only on the variance of $\chi$ and on the parameter
$\nu_3$ quantifying the amount of non-Gaussianity. The variance is
directly proportional to the value of the Hubble constant at horizon
crossing, $H_c$. The parameter $\nu_3$ is proportional to $\lambda$
and is explicitly given by
\begin{equation}
\nu_3=-{\lambda\,N_e}/({3\,H_c^2})\approx -0.60.
\end{equation}
On Fig.~\ref{chipdf}, the numerical results are compared to the
expression (\ref{pdfapprox}) and to a Gaussian distribution of the
same variance. It confirms that the analytic form
(\ref{pdfapprox}) is a very good approximation of the PDF. The
value of $\nu_3$ in this plot is~\footnote{Note that the variance
of the field value evolves due to the non-zero effective mass of
$\chi$.} $\nu_3=-{0.15}/{\sigma_{\chi}^2}$, which implies that the
value of the kurtosis of distribution is
\begin{equation}\label{kurtosis}
  s_4\equiv{\langle\chi^4\rangle}/{\langle\chi^2\rangle^2}-3=4\nu_3\sigma_{\chi}^2.
\end{equation}
The curvature fluctuation PDF is related to the PDF of the field
$\chi$. In the limit where the inflation ends abruptly the
relation is straightforward, Eq.~(\ref{chiinducedR}). It is more
complex in general and depends on the details of the phase of
inflation.

The parameters of the potential (\ref{threefieldmodel}) are tuned,
as usual, so that the amplitude of the fluctuations are compatible
with the observations. No further fine-tunings are required for
the initial conditions: whatever the initial value for the field
$\chi$, it rolls down rapidly towards $\chi=0$ to follow the
valley bottom, as described in the previous paragraphs.

The model (\ref{threefieldmodel}) reads as a simple polynomial
potential of order four but involves three fields. It is actually
possible to build models involving only two fields. In this case, it is
clear that the $\chi$ trajectory cannot be straight and has to bend. A
possible explicit model can be obtained when the zero-mode trajectory
is given by
\begin{equation}\label{chitraj}
  \chi_0(t)=\chi_{\infty}\,\tanh\left(\frac{\alpha\,
  \phi_0(t)}{\chi_{\infty}}\right)
\end{equation}
where $\alpha$ and $\chi_{\infty}$ are free parameters. A model to
get such a trajectory can be constructed by considering a term of
the form $(\chi-\chi_0)^4$ in the potential, as first advocated in
Ref.~\cite{ub02}, so that the potential might take  the form
\begin{equation}\label{twofieldmodel0}
  V(\phi,\chi)=\frac{1}{2}m^2\,\phi^2+\frac{\lambda}{4!}\left(
  \chi\cosh\left[\alpha\,\phi\over\chi_{\infty}\right]-
  \chi_{\infty}\sinh\left[\alpha\,\phi\over\chi_{\infty}\right]\right)^4.
\end{equation}
Although it gives the adequate trajectory (\ref{chitraj}) in field
space, it is non-polynomial and does not induce significant metric
non-Gaussianities. The reason is that the $\chi$ fluctuations are
squashed as soon as the trajectory is bent, the bundle of
trajectories being drastically focused much before any significant
bending. This is a generic features of all models in which the
bending of the trajectory is driven by the non-linear coupling
term.

\begin{figure*}
\centerline{\begin{tabular}{cc}
\psfig{figure=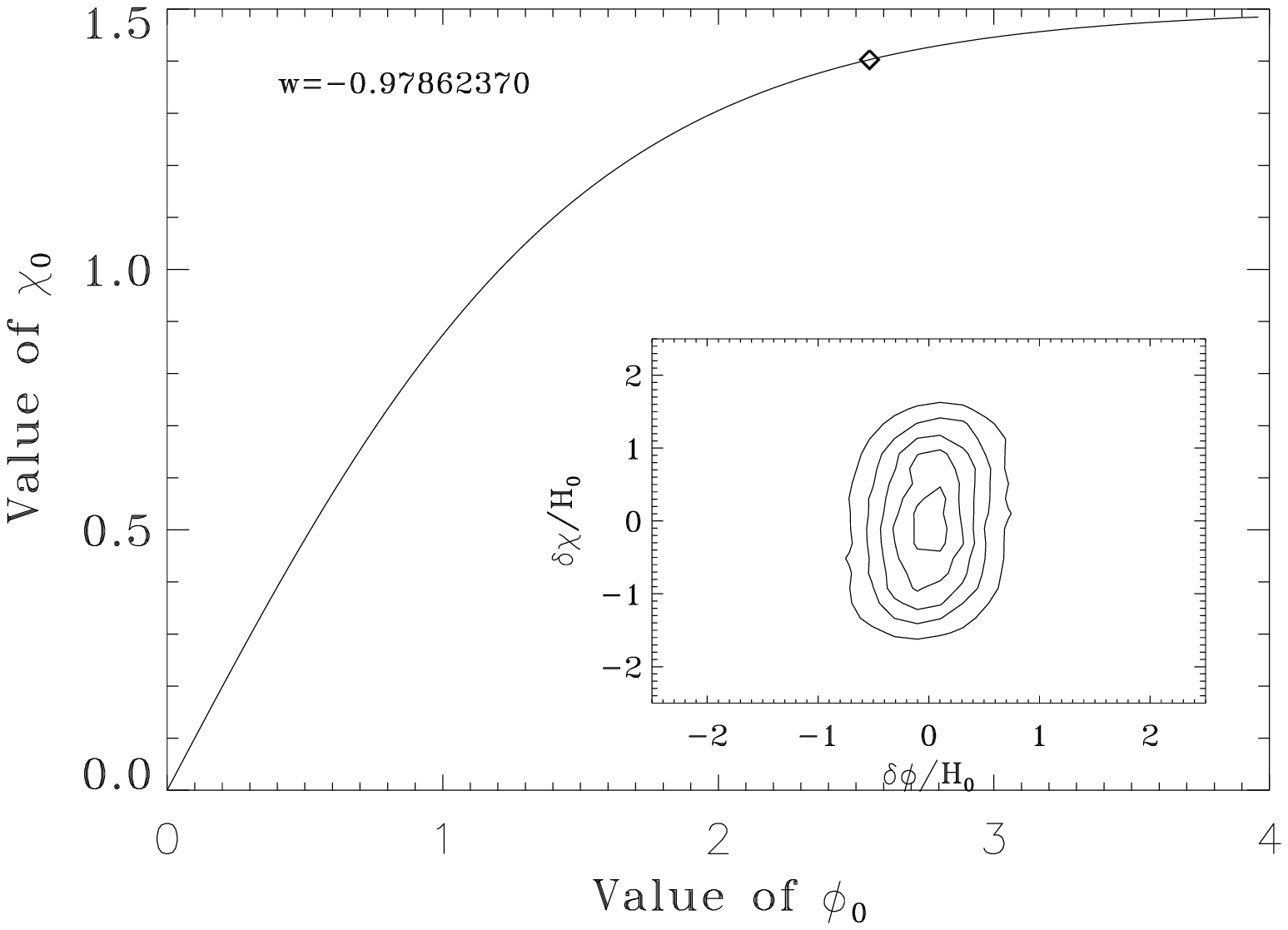,width=7cm} &
\psfig{figure=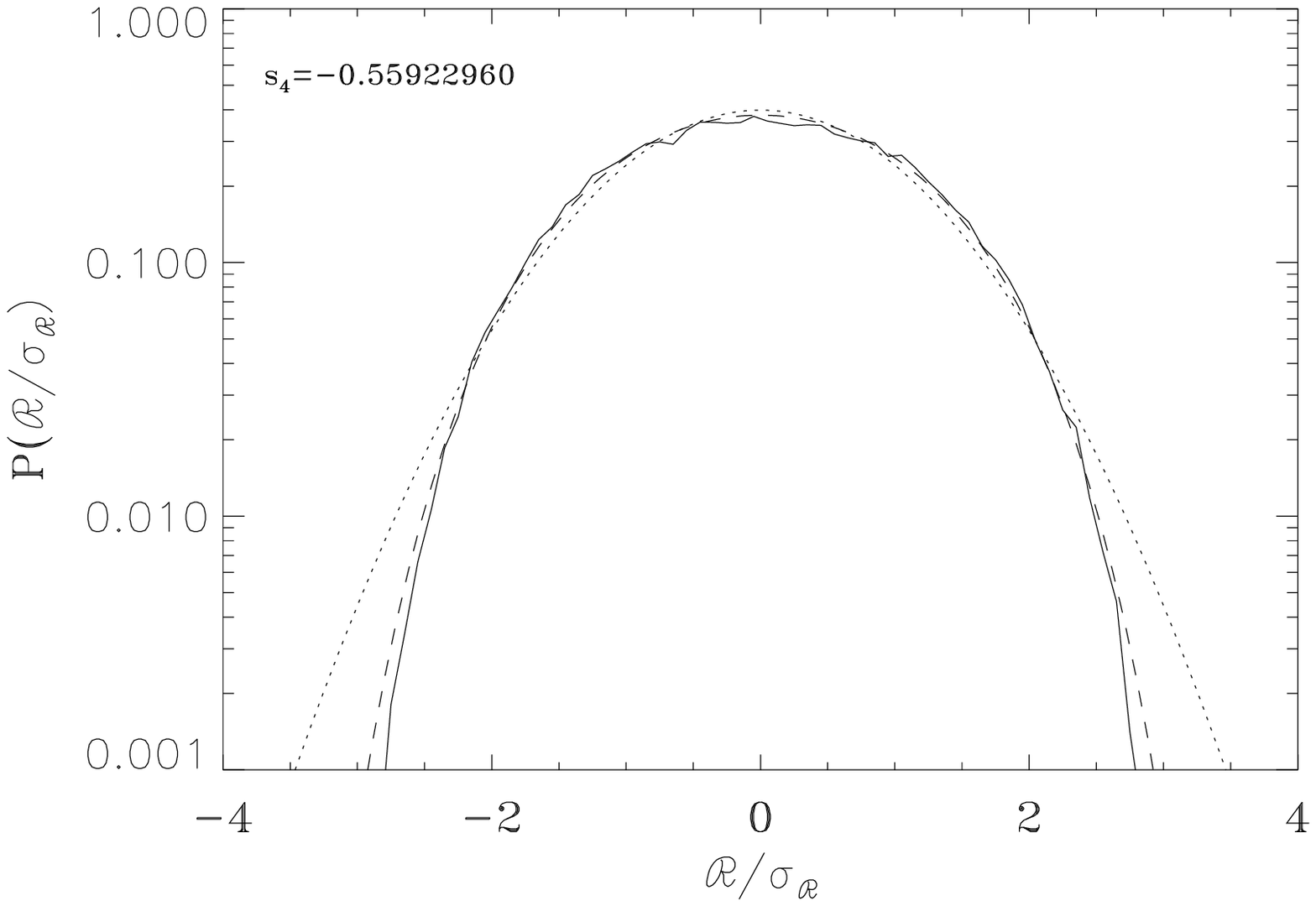,width=7cm} \\
\psfig{figure=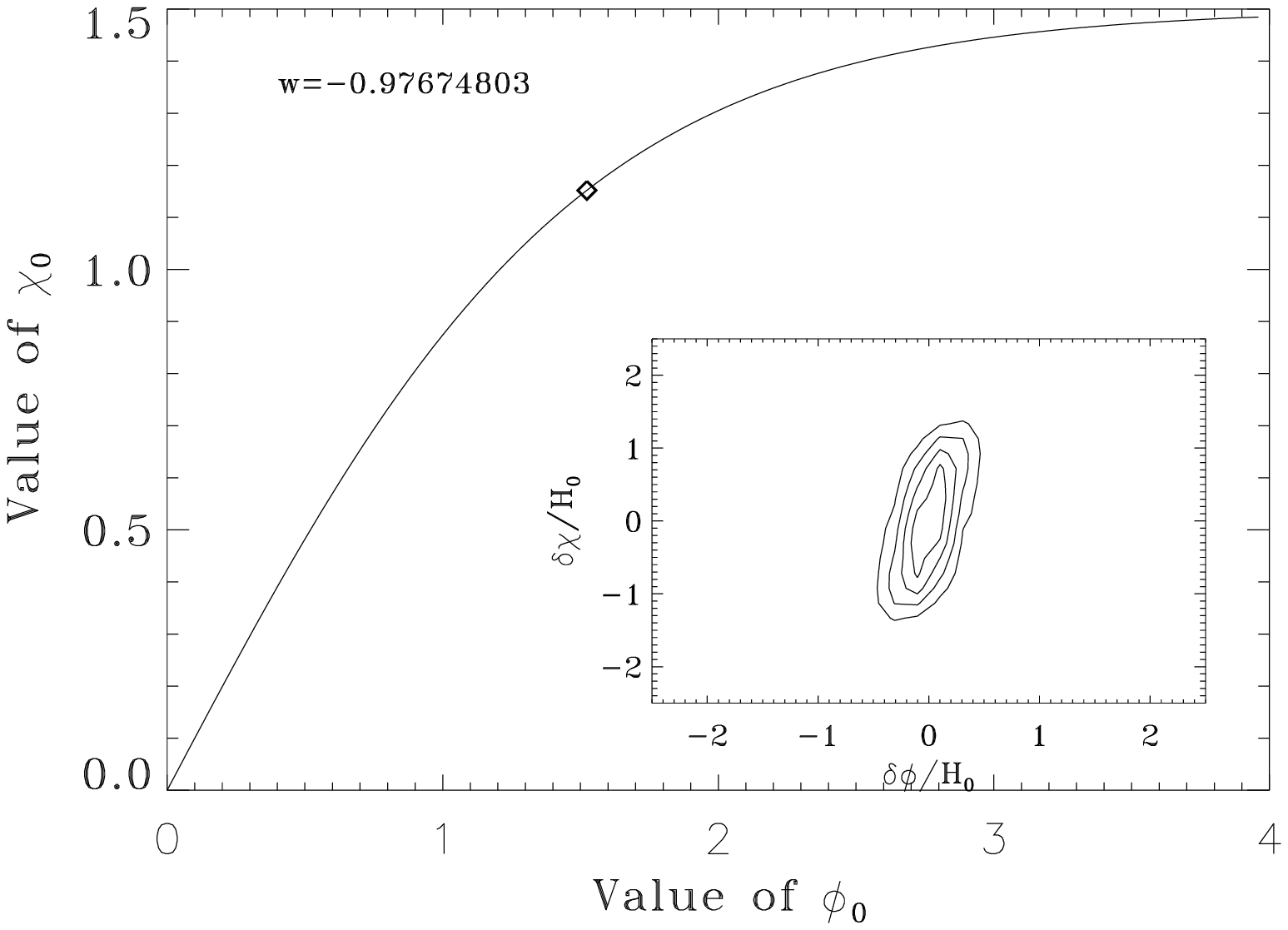,width=7cm} &
\psfig{figure=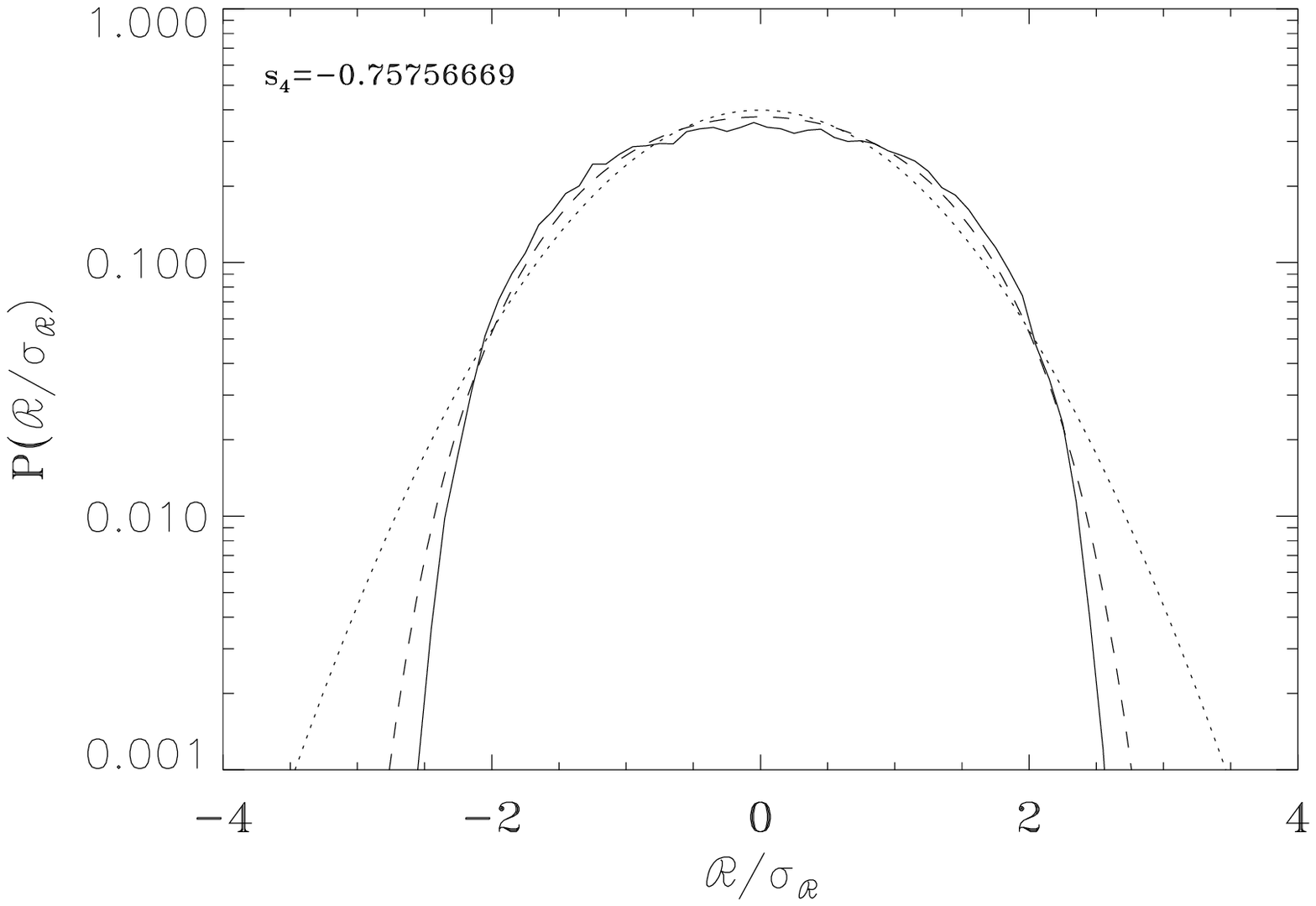,width=7cm} \\
\psfig{figure=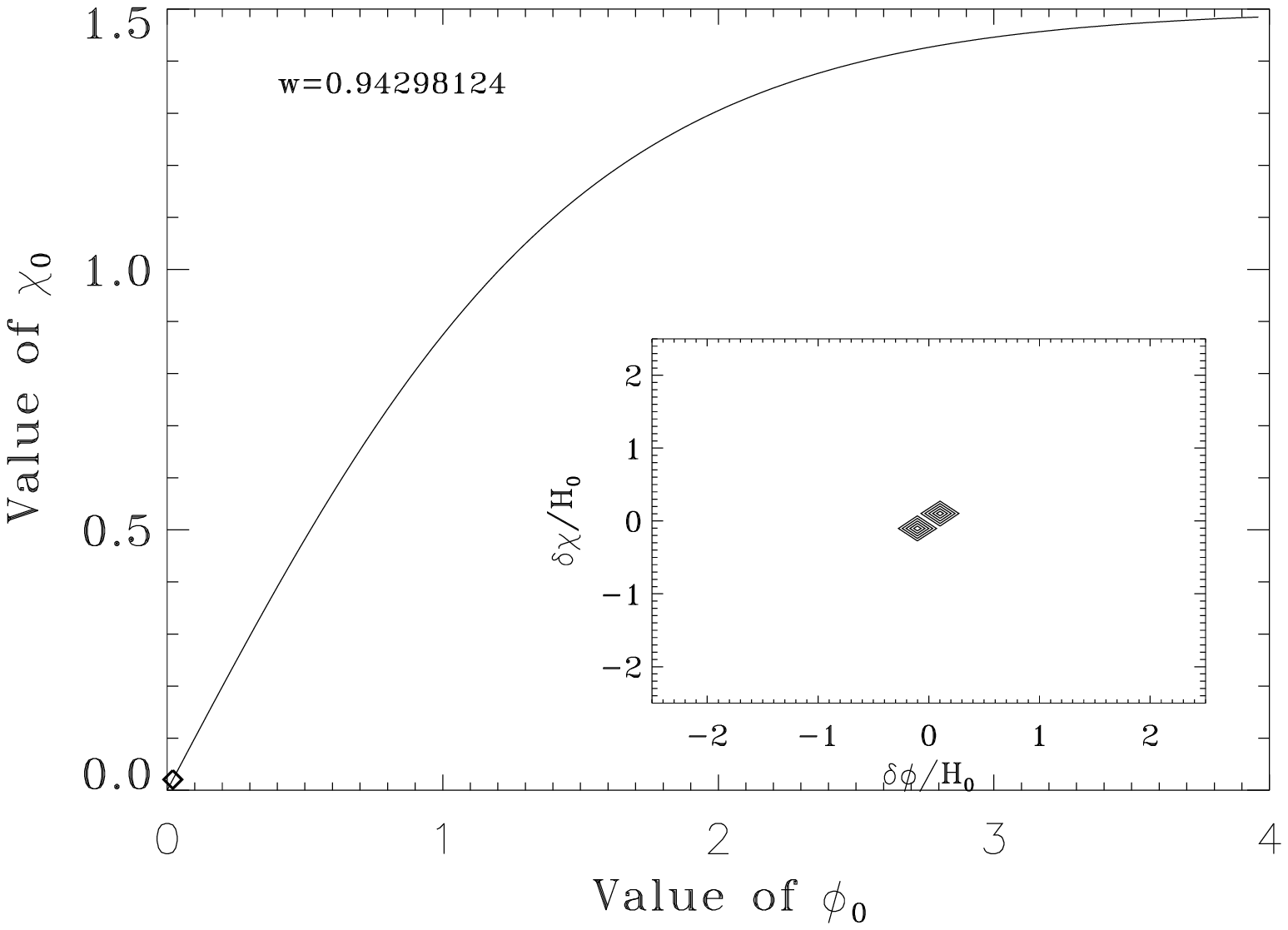,width=7cm} &
\psfig{figure=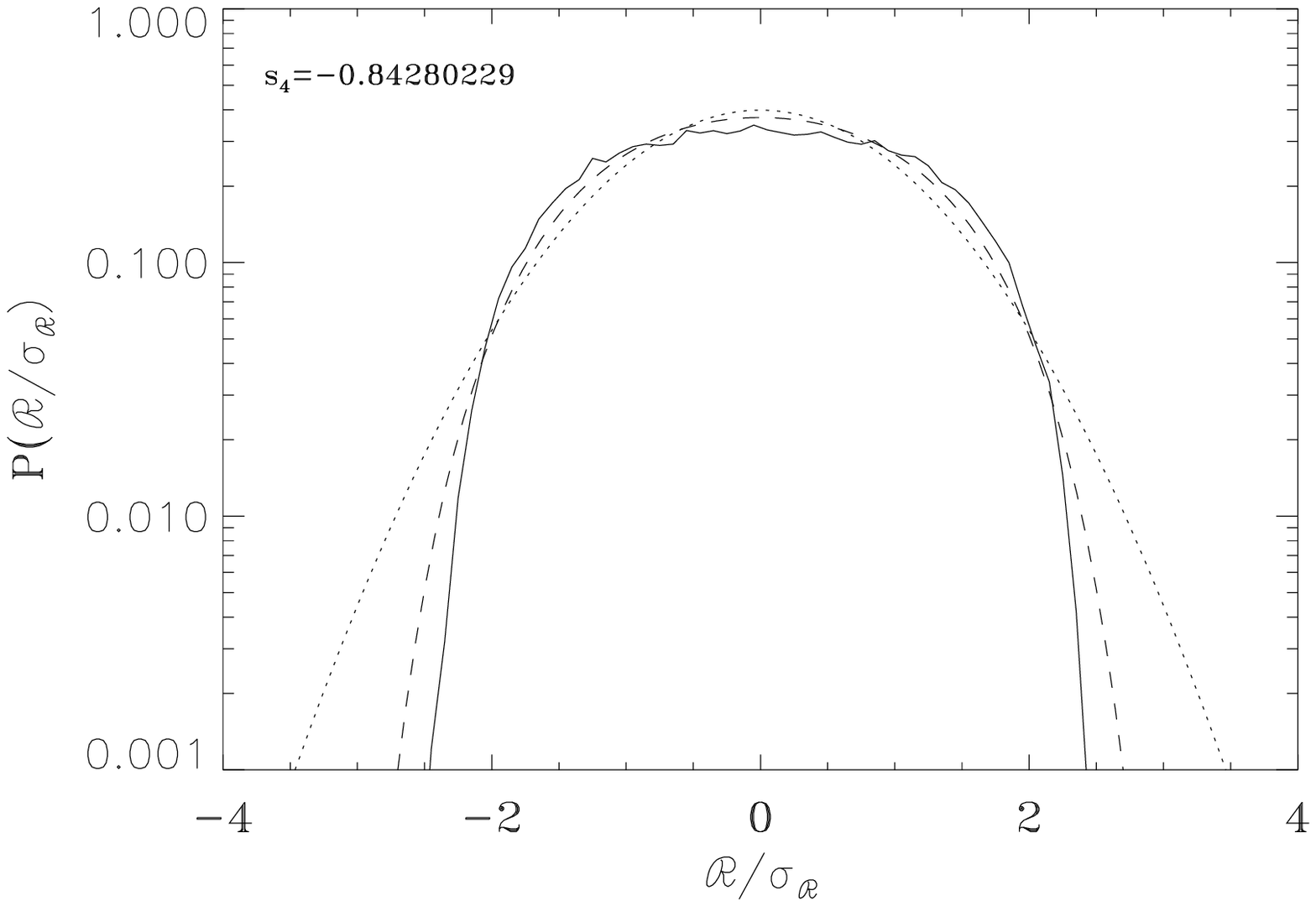,width=7cm} \\
  \end{tabular}
 }
  \caption{PDF for the two-field model with potential (\ref{twofieldmodel1})
  at different timestep with the parameters (\ref{para1}-\ref{parafin}).
  The left panel describes the field trajectory and the shape of
  the wavepacket and the right panels compare the numerically obtained
  PDF (solid) to a Gaussian (dotted) and to the analytical PDF (\ref{pdfapprox})
  (dash). $w$ is the effective equation of state parameter and $s_4$ is defined in
  Eq.~(\ref{kurtosis}).}
  \label{chi4pdf_mod1}
\end{figure*}

A way to construct a model in which the $\chi$ fluctuations are
not squashed is to consider the potential
\begin{eqnarray}\label{twofieldmodel1}
  V(\phi,\chi)&=&\frac{1}{2}m^2\left(\chi^2+\frac{\chi^2_{\infty}}{\alpha^2}
  \sinh^2\left[\alpha\,\phi\over\chi_{\infty}\right]\right)+\nonumber\\
  &&\frac{\lambda}{4!}\left(
  \chi\cosh\left[\alpha\,\phi\over\chi_{\infty}\right]-
  \chi_{\infty}\sinh\left[\alpha\,\phi\over\chi_{\infty}\right]\right)^4.
\end{eqnarray}
Now, the zero mode trajectory $(\phi_0,\chi_0)$ is not determined
by the non-linear coupling. At lowest order in the coupling
parameter $\lambda$ and in the slow-roll regime, the Klein-Gordon
equations for $\phi$ and $\chi$ can be integrated and lead to the
trajectory (\ref{chitraj}).

In Fig.~\ref{chi4pdf_mod1}, we depict the behavior of the
super-horizon $\phi$ and $\chi$ fluctuations (left panels) and the
curvature fluctuations that are induced by the initial $\chi$
fluctuations (right panels). The parameters of the model for the
simulation of Fig.~\ref{chi4pdf_mod1} were chosen to be
\begin{eqnarray}\label{twofieldparameters}
  &&m=10^{-7.}\,M_{\rm Pl.}\qquad \alpha=1 \label{para1}\\
  &&\chi_{\infty}=1.5\,M_{\rm Pl.}\qquad
  \lambda=5\times10^{-5},\label{parafin}
\end{eqnarray}
so that it gives realistic fluctuation amplitudes and e-fold
number. The initial conditions are set up at $\phi_{\rm
init}=4M_{\rm Pl.}$. At this time the coupling constant is then
about $0.15$ and the number of $e$-folds during the inflationary
period is about $72$. We integrate the Klein-Gordon equations and
the Friedmann equation for the set of classical trajectories
starting from different initial conditions around $(\phi_{\rm
init},0)$ Gaussian distributed with $\left<(\phi-\phi_{\rm
init})^2\right>\sim\left<\chi^2\right>\sim H^2_{\rm init}$.

As can be observed on the left panels the joint evolution of the
$\phi$ and $\chi$ fluctuations is quite complex. This is due to
the fact that the fluctuation distribution is shaped by the
nonlinear couplings during the trajectory bend. The resulting
curvature fluctuations can however be simply described. They are
given by the sum of those induced by the $\phi$ fluctuations,
$\mR_{\phi}$, (Gaussian distributed) and those induced by the
$\chi$ fluctuations, $\mR_{\chi}$. Both distributions have
approximately the same variance. For the parameters
(\ref{para1}-\ref{parafin}), our numerical computations show that
actually the r.m.s. of the former is about $18\,H_c$, the r.m.s.
of the latter $15\,H_c$. The right panels show the PDF shape of
the non-Gaussian component of the curvature fluctuations,
$\mR_{\chi}$. They follow the distribution function given in
Eq.~(\ref{pdfapprox}) as long as $\vert s_4\vert \lesssim 0.7 $ at
least for less than $3\sigma$ events. The complete PDF of the
curvature fluctuations can then be obtained as the convolution of
the PDFs of the two components, $\mR_{\phi}$ and $\mR_{\chi}$.

In this {\it letter}, we have introduced a series of multi-field
inflationary models leading to the generation of non-Gaussian
primordial curvature fluctuations. We have been able to build
explicit models with either two or three scalar fields. In the
case of two-field models, it is difficult to design a simple
potential: it has to be constructed in such a way that the
non-linear term does not damp the curvature fluctuations. In the
case of three fields, we have shown that models as simple as
polynomial potentials of order four work. In both cases we have
illustrated our models by numerical computations and have shown
that the analytic result (\ref{pdfapprox}) derived in
Ref.~\cite{ub02} is a very good approximation to the primordial
curvature fluctuation PDF when non-Gaussianities in the $\chi$
distribution remain modest. Let us emphasize that in the two-field
models the potential has to be tuned but that this is not the case
for the three-field model. Moreover, the initial conditions do not
require to be fine-tuned, as explained before. This description of
the primordial non-Gaussianity is the starting point of further
investigations on its various observational aspects, both for CMB
and large scale structure surveys~\cite{preparation}.

\end{document}